# Phase equilibria, volumetric properties, viscosity, and thermal conductivity of hydrogen binary mixtures containing methane, nitrogen, carbon monoxide, and carbon dioxide

Ehsan Heidaryan, Saman A. Aryana[*]

*Department of Chemical & Biomedical Engineering, University of Wyoming, Laramie, WY, 82071, USA*

**Abstract**

This works proposes modifications to the values of input parameters in widely accepted semi-analytical models of thermodynamic properties of hydrogen and its binary mixtures to improve their prediction accuracy. The modifications include characterization parameters for the Peng and Robinson Equation of State (PR-EoS), the Jossi-Stiel-Thodos (JST)/Lohrenz-Bray-Clark (LBC) viscosity correlation, and the Chung-Lee-Starling (CLS) thermal conductivity correlation, applicable to pure hydrogen and its binary mixtures containing methane, nitrogen, carbon monoxide, and carbon dioxide. The changes preserve the mathematical structure of the models and ensure their compatibility with existing commercial packages. We consider temperatures ranging from 90 K (triple point of methane) to 1000 K, and pressures ranging from atmospheric up to 100 MPa. The revised characterization parameters of hydrogen lead to a reduction in the Absolute Average Deviation (AAD) of the PR-EoS calculation for hydrogen density from 2.76% (with a maximum of 8.05%) down to 0.182% (with a maximum of 1.9%). For the JST/LBC viscosity correlation, the AAD decreases from 11.82% (with a maximum of 54.3%) to 4.08% (with a maximum of 13%). Regarding thermal conductivity, the adjustments for the CLS correlation results in the AAD decreasing from 36% (with a maximum of 44%) to 2.14% (with a maximum of 12.1%). Binary interaction parameters for gaseous mixtures containing hydrogen and phase equilibria calculation are also proposed. Finally, we investigage optimal mixing rules parameters for viscosity and thermal conductivity of mixtures for more accurate predictions of mixtures' transport properties.

[*] Corresponding author: E-mail address: saryana@uwyo.edu (S.A. Aryana).

# 1. Introduction

Hydrogen ($H_2$) is often heralded as a key energy carrier in energy systems of the future [1-2]. The impetus for such a transition from carbon based fuels to hydrogen is its potential to produce energy without emitting pollutants and Greenhouse Gases (GHGs) [3], thereby reducing the world's heavy reliance on fossil fuels and the associated environmental degradation [4].

In the transition to a hydrogen-dominant energy landscape, accurate and reliable estimates of thermodynamic and transport properties of $H_2$ and its mixtures is central to its effective implementation in the energy system. These properties influence how $H_2$ is produced, stored, transported, and utilized, so understanding them is essential for the safe and efficient integration of $H_2$ and related mixtures into the global energy landscape.

Given hydrogen's critical temperature (33.19 K) and pressure (1.293 MPa), it predominantly exists in the gaseous phase, particularly in the context of most of its anticipated energy applications. Accurate density calculations are crucial for calculating the energy content of $H_2$, whether in storage or in transit, and are pivotal in determining storage capacities, ensuring that facilities meet stipulated specifications and operate at peak efficiency. Furthermore, reliable and accurate estimates of hydrogen's properties and energy content play a critical role in safety and risk assessments. In commercial operations, accurate volumetric and density calculations influence infrastructure design and legal and contractual obligations. Other examples of the practical significance of accuerate amd reliable thermodynamic properties of hydrogen and its mixtures include thermal management and convective heat transfer in power generation, reaction spontaneity and safety procedures, flow and pressure for transport and pipelines systems, and the prediction of the the energy either released or absorbed during industrial processes involving hydrogen.

As such, we require relable and accurate models, i.e., equations of state (EoS), viscosity and thermal conductivity correlations. Koo et al. [5] conducted a study on integrating $H_2$ into high-pressure natural gas pipelines, highlighting the vital role of thermophysical properties in capturing the real behavior of gases. They underscored the necessity of using an EoS to accurately estimate the fluid hydraulics in pipelines. Nova et al. [6] stressed the importance of understanding the thermodynamic behavior of systems involved in $H_2$ production, and the challenges arising from hydrogen sulfide ($H_2S$) emissions. Liu et al. [7] emphasized the need for accurate calculations in estimating the density, viscosity, and thermal conductivity for the decompression wave speed curve in a blend of natural gas and $H_2$ (NGH2), as accurate estaimtes are crucial for determining arrest toughness. Tan et al. [8] explored energy costs associated with transporting mixtures of $H_2$ and methane (Hythane) via the Energy Specific Toll (EST) value and showcasing the pivotal role of thermodynamics in ensuring efficient and safe Hythane transportation. Jia et al. [9] investigated the challenges associated with understanding the behavior of Hythane, particularly concerning leakage and diffusion in compressor plants. Li et al. [10] studied the behavior of Hythane in closed environments, underscoring the central role of thermodynamics in its energy and flow dynamics. Bellis et al. [11] demonstrated the importance of thermodynamics in studying combustion processes, while Chen et al. [12] illustrated how transport properties are critical for accurate simulations and insights into combustion behavior. Building on work by Fang et al. [13] and Naquash et al. [14], thermophysical models have been established for accurate heat transfer and flow friction calculations, particularly in scenarios involving large temperature drops and high heat fluxes. Lee et al. [15] investigated the reaction rate constant of $H_2$ hydrate formation for $H_2$ storage, highlighting the significant impact of the EoS choice on the process's speed and efficiency, relevant to $H_2$ storage and retrieval. Li et al. [16] analyzed the Joule-Thomson effect in a high-pressure hydrogen pressure-reducing valves using various EoSs, revealing the substantial influence of the EoS selection on the outcome of the modeled throttling process. Chai et al. [17] studied the complex interactions of $H_2$ with underground rocks and aquifer fluids, incorporating multiple injection/withdrawal cycles and considering cushion gas to enhance $H_2$ round-trip efficiency. Their results indicated that accurate calculations of thermophysical properties are crucial for understanding residual trapping and hydrogen production. Shoushtari et al. [18] also underscored the importance of thermophysical properties in underground hydrogen storage (UHS). In their thorough analysis of green $H_2$ production, Al-Mahgari et al. [19] utilized an EoS for accurate modeling of the process, highlighting its critical role in calculating pre-electrolysis compression and heating under high temperature and pressure conditions. According to Hematpur and colleagues [21], the thermal conductivity of $H_2$ plays a crucial role in determining changes in pressure and temperature during storage operations.

This work introduces a series of methodological amendments to widely accepted semi-analytical methods, including the Peng and Robinson Equation of State (PR-EoS) [21], the Jossi-Stiel-Thodos (JST) [22] (generalized by Lohrenz-Bray-Clark (LBC) [23]) viscosity correlation, and the Chung-Lee-Starling (CLS) [24] thermal conductivity correlation (simplified by Elliott et al. [25]). These refined methodologies apply to both pure hydrogen and its mixtures, aiming to enhance the accuracy and comprehensiveness of our understanding of their thermodynamic properties. This study is focused on a temperature range starting at the triple-point of methane, with pressures extending up to 100 MPa.

## 2. Thermophysical models

There are a several models in the literature that describe the thermodynamics [26], viscosity [27], and thermal conductivity [28] of $H_2$. However, accurately modeling hydrogen mixtures remains a challenge, especially since such mixtures are unavoidable in many industrial applications, including the energy sector. In these scenarios, Cubic Equations of State (CEoS) become indispensable due their relatively accurate predictions of mixture behavior under various physical conditions — a crucial aspect for many engineering applications. Additionally, universal models of the viscosity and thermal conductivity of mixtures are critical engineering tools.

*2.1. Peng-Robinson Equation of State (PR-EoS/pρT model)*

Since its inception in 1976, the Peng-Robinson Equation of State (PR-EoS) [21] has proven to be accurate in predicting the behavior of fluids, particularly in chemical and petroleum industries. Due to its simplicity and reliability, PR-EoS has been widely used in both academic research and industry for calculating phase equilibria and thermodynamic properties of mixtures. The PR-EoS model is defined by

$$p = \frac{RT}{v-b} - \frac{a}{v^2 + 2vb - b^2}, \tag{1}$$

where $a$ and $b$ are the EoS parameters. The pure component parameters ($a$ and $b$) can be calculated as

$$\begin{aligned} a &= \Omega_a \frac{R^2 T_c^2}{p_c} \alpha(T) \quad \& \quad \alpha(T) = \left(1 + m\left(1 - \sqrt{T/T_c}\right)\right)^2 \\ b &= \Omega_b \frac{RT_c}{p_c} \end{aligned} \tag{2}$$

In Equation 2, $\Omega_a$, $\Omega_b$, and $m$ can be calculated as

$$\begin{aligned} \Omega_a &= \frac{8(5X+1)}{49-37X} \approx 0.45724... \\ \Omega_b &= \frac{X}{X+3} \approx 0.07780... \\ X &= \left(-1 + \sqrt[3]{6\sqrt{2}+8} - \sqrt[3]{6\sqrt{2}-8}\right)/3 \\ m(\omega) &= \begin{cases} 0.37464 + 1.54226\omega - 0.26992\omega^2, & \omega \leq 0.49 \\ 0.3796 + 1485\omega - 0.1644\omega^2 + 0.01667\omega^3, & \omega > 0.49 \end{cases} \end{aligned} \tag{3}$$

In PR-EoS [21], compound properties of the critical pressure ($p_c$), critical temperature ($T_c$), and acentric factor ($\omega$) are used as inputs for each fluid characterization. The PR-EoS [21] can be extended to mixtures of fluids using mixing rules, as

$$a_{mix} = \sum_{i=1}^{n}\sum_{j=1}^{n} y_i y_j \sqrt{a_i a_j}\left(1-k_{ij}\right)$$
$$b_{mix} = \sum_{i=1}^{n} y_i b_i$$
(4)

where $y_i$ and $y_j$ are the mole fraction of components $i$ and $j$ in the mixture, and $a_i$, $a_j$, and $b_i$ are pure components' $a$ and $b$ parameters. The binary interaction parameter $k_{ij}$ accounts for the deviation from ideal behavior when two different substances are mixed (typically determined by fitting experimental data).

### 2.2. Lohrenz et al. viscosity correlation (LBC ρμT model)

A set of correlations was developed by Jossi et al. [22] for fourteen pure substances in dense gaseous and liquid phases. They derived four different quartic polynomials of reduced density ($\rho/\rho_c$) for water, ammonia, hydrogen, and normally behaving substances. Lohrenz et al. [23] adapted this work by Jossi et al. [22] to estimate the viscosity of reservoir fluids (the LBC correlation). The LBC correlation is known for its simplicity and accuracy over a wide range of temperatures and pressures, making it widely used in the petroleum and chemical industries. The LBC correlation is expressed as

$$\left((\mu - \mu^*)\xi + 10^{-4}\right)^{1/4} =$$
$$0.1023 + 0.023364\left(\frac{\rho}{\rho_c}\right) + 0.058533\left(\frac{\rho}{\rho_c}\right)^2$$
$$-0.040758\left(\frac{\rho}{\rho_c}\right)^3 + 0.0093324\left(\frac{\rho}{\rho_c}\right)^4$$
(5)

where $\xi$ is the viscosity-reducing parameter and is defined by

$$\xi = T_c^{1/6} M^{-1/2} p_c^{-2/3}.$$
(6)

and $\mu^*\xi$ is the dilute gas viscosity [29], which can be calculated using,

$$\mu^*\xi = \begin{cases} 34\times 10^{-5}\left(\frac{T}{T_c}\right)^{0.94}, & \frac{T}{T_c} \leq 1.5 \\ 17.78\times 10^{-5}\left(4.58\left(\frac{T}{T_c}\right)-1.67\right)^{5/8}, & \frac{T}{T_c} > 1.5 \end{cases}.$$
(7)

For mixtures, the following combination rules may be used,

$$\rho_{c,mix} = \frac{1}{\sum_{i}^{n} y_i V_{c_i}}$$

$$\xi_{mix} = \left(\sum_{i=1}^{n} y_i T_c\right)^{1/6} \left(\sum_{i=1}^{n} y_i M\right)^{-1/2} \left(\sum_{i=1}^{n} y_i p_c\right)^{-2/3}.$$

$$\mu_{mix}^* = \sum_{i=1}^{n} y_i \mu_i^* \sqrt{M_i} \bigg/ \sum_{i=1}^{n} y_i \sqrt{M_i} \qquad (8)$$

The dilute gas viscosity combination rule has been proposed by Herning and Zipperer [30]. The pseudo-reduced temperature ($T/\Sigma y_i T_{ci}$) is incorporated into Equation 7 for mixtures. When utilizing this correlation, it is essential to note that the value of $\rho/\rho_c$ should not exceed 3.

*2.3. Chung et al. thermal conductivity correlation (CLS ρλT model)*

Chung et al. [24], building upon the foundational work presented in Chung et al. [31], meticulously extended the thermal conductivity model initially developed for low-pressure gases, adapting it to encompass fluids under conditions of elevated densities. This adaptation was achieved by incorporating empirically derived, density-dependent functions, showcasing a commendable effort in bridging the gap between low-pressure gas behaviors and the complex dynamics of dense fluids. In the development of these correlations, Pitzer's acentric factor ($\omega$), the dimensionless dipole moment ($\mu_r$), and an empirically ascertained association parameter ($\kappa$) have been used. These pivotal parameters are instrumental in encapsulating the multifaceted influences exerted by the molecular structure of polyatomic molecules, the polar effects, and the hydrogen-bonding interactions, respectively. The nuanced characterization facilitated by these parameters culminates in a robust framework, enabling the accurate prediction of thermal conductivity in polar fluids. This method is simplified by Elliott et al. [25] and can be written as

$$\lambda = \frac{31.2\mu^0\Psi}{M'}\left(G_2^{-1} + yB_6\right) + qB_7G_2y^2\sqrt{T/T_c}$$

$$\mu^0 = 40.785\frac{F_c\sqrt{MT}}{V_c^{2/3}\Omega_v}$$

$$F_c = 1 - 0.2756\omega + 0.059036\mu_r^4 + \kappa$$

$$\Omega_v = 1.16145(T^*)^{-0.14874} + 0.52487\left(Exp(-0.7732T^*)\right)$$
$$+ 2.16178\left(Exp(-2.43787T^*)\right)$$

$$T^* = 1.2593(T/T_c)$$

$$\Psi = 1 + \frac{0.215\alpha + 028288\alpha^2 - 1.061\alpha\beta + 0.6665\alpha\vartheta}{0.6366 + \beta\vartheta + 1.061\alpha\beta} \qquad (9)$$

$$\alpha = \frac{C_p^0 - R}{R} - \frac{3}{2}$$

$$\beta = 0.7862 - 0.7109\omega + 1.3168\omega^2$$

$$\vartheta = 2 + 10.5\left(\frac{T}{T_c}\right)^2$$

$$y = \frac{V_c}{6V}$$

$$G_1 = \frac{1 - y/2}{(1-y)^3}$$

$$G_2 = \frac{(B_1/y)(1 - Exp(-B_4 y)) + B_2G_1Exp(B_5 y) + B_3G_1}{B_1B_4 + B_2 + B_3}$$

$$q = 3.586 \times 10^{-3}\frac{\sqrt{T_cM'}}{V_c^{2/3}}$$

$$B_i = a_i + b_i\omega + c_i\mu_r^4 + d_i\kappa$$

where $M' = M/10^3$, and the constants for $B_i$ are listed in Table 1.

Table 1. Constants for the generalized thermal conductivity correlation proposed by Chung et al. [24] (simplified by Elliott et al. [25]).

| $i$ | $a_i$    | $b_i$    | $c_i$    | $d_i$   |
|-----|----------|----------|----------|---------|
| 1   | 2.4166   | 0.74824  | -0.91858 | 121.72  |
| 2   | -0.50924 | -1.5094  | -49.991  | 69.983  |
| 3   | 6.6107   | 5.6207   | 64.76    | 27.039  |
| 4   | 14.543   | -8.9139  | -5.6379  | 74.344  |
| 5   | 0.79274  | 0.82019  | -0.69369 | 6.3173  |
| 6   | -5.8634  | 12.801   | 9.5893   | 65.529  |
| 7   | 91.089   | 128.11   | -54.217  | 523.81  |

Adopting the three-parameter corresponding state method as described by Lee et al. [32], mixing rules have been established to apply this method for estimating the thermal conductivity of high-pressure gas mixtures as

$$T^*_{mix} = T/(\varepsilon/k_B)_{mix}$$
$$T_{c_{mix}} = 1.2593(\varepsilon/k_B)_{mix}$$
$$(\varepsilon/k_B)_{mix} = \sum_i^n \sum_j^n y_i y_j (\varepsilon_{ij}/k_B) \sigma_{ij}^3 / \sigma_{mix}^3$$
$$\sigma_{mix}^3 = \sum_i^n \sum_j^n y_i y_j \sigma_{ij}^3$$
$$V_{c_{mix}} = (\sigma_{mix}/0.809)^3$$
$$M_{mix} = \sum_i^n \sum_j^n y_i y_j (\varepsilon_{ij}/k_B) \sigma_{ij}^2 M_{ij}^{1/2} / (\varepsilon/k_B)_{mix} \sigma_{mix}^2$$
$$\omega_{mix} = \sum_i^n \sum_j^n y_i y_j \omega_{ij} \sigma_{ij}^3 / \sigma_{mix}^3$$
$$\mu_{mix}^4 = \sigma_{mix}^3 \sum_i^n \sum_j^n (y_i y_j \mu_i^2 \mu_j^2 / \sigma_{ij}^3)$$
$$\mu_{r_{mix}} = 131.3 \mu_{mix} / (V_{c_{mix}} T_{c_{mix}})^{1/2}$$
$$\kappa_{mix} = \sum_i^n \sum_j^n y_i y_j \kappa_{ij} \qquad , \qquad (10)$$
$$C^0_{p_{mix}} = \sum_{i=1}^n y_i C^0_{p_i}$$

where the binary parameters in this context are determined using the following combining rules

$$\sigma_{ij} = \xi_{ij} (\sigma_i \sigma_j)^{1/2}$$
$$\sigma_{ii} = \sigma_i = 0.809 V_{c_i}^{1/3}$$
$$\varepsilon_{ij}/k_B = \zeta_{ij} ((\varepsilon_i/k_B)(\varepsilon_j/k_B))^{1/2}$$
$$\varepsilon_{ii}/k_B = \varepsilon_i/k_B = T_{c_i}/1.2593 \qquad ,$$
$$\omega_{ij} = (\omega_i + \omega_j)/2 \quad \& \quad \omega_{ii} = \omega_i \qquad (11)$$
$$\kappa_{ij} = (\kappa_i \kappa_j)^{1/2} \quad \& \quad \kappa_{ii} = \kappa_i$$
$$M_{ij} = 2 M_i M_j / (M_i + M_j)$$

where $\xi_{ij}$ and $\zeta_{ij}$ represent binary interaction parameters, which are typically set equal to unity.

## 2.4. Ideal gas isobaric heat capacity ($C_p^0$)

As Bartolomeu and Franco [33] discussed, the isobaric heat capacity ($C_p$) of H2 at temperatures above 100K is predominantly dictated by the properties of an ideal gas isobaric heat capacity. The equation is grounded by Aly and Lee [34] in physical meanings derived from statistical mechanical derivation, providing a robust theoretical basis for the calculations as

$$C_p^0 = d_0 + d_1 \left( \frac{d_2/T}{\text{Sinh}(d_2/T)} \right)^2 + d_3 \left( \frac{d_4/T}{\text{Cosh}(d_4/T)} \right)^2. \tag{12}$$

Considering SI units, in this equation, $T$ is in Kelvin (K), and the ideal gas isobaric heat capacity ($C_p^0$) is in Joules per mole-Kelvin (J/mole-K). Using pressure with an exponent improvement by Heidaryan and Aryana [35], other ideal gas properties can be calculated based on this equation. The ideal gas enthalpy can be calculated as

$$H^0 = d_0 T + d_1 T (d_2/T) \text{Coth}(d_2/T) - d_3 T (d_4/T) \text{Tanh}(d_4/T) + d_5, \tag{13}$$

and the ideal gas entropy as

$$S^0 = d_0 Ln(T) + d_1 \left( (d_2/T) \text{Coth}(d_2/T) - ln(\text{Sinh}(d_2/T)) \right) \\ - d_3 \left( (d_4/T) \text{Tanh}(d_4/T) - ln(\text{Cosh}(d_4/T)) \right) - d_6 ln^{d_7}(p) \tag{14}$$

The abovementioned computational methodologies are straightforward, making them accessibile and expanding their utility across a wide range of scientific and engineering applications [36—40].

## 3. Applied amendments for pure hydrogen

For pure hydrogen, this work uses 1407 density points, 910 viscosity points, and 2265 thermal conductivity points from the data bank selected by Heidaryan and Aryana [35]. Using the Nelder and Mead [41] simplex method, the input parameter for each method is determined and listed in Table 2.

Table 2. Fitted parameter for pure $H_2$.

| PR-EoS $p\rho T$ model (Eq.1) | LBC-viscosity $\rho\mu T$ model (Eq.5) | CLS-Thermal Conductivity $\rho\lambda T$ model (Eq.9) | $C_p^0$, $H^0$ and $S^0$ (Eq.12, 13 and 14) |
|---|---|---|---|
| $T_c$=32.06 [K] | $T_c$=19.47 [K] | $T_c$=145.39 [K] | $D_0$=22.75612060 [J/mol-K] |
| $p_c$=1269018 [Pa] | $p_c$=1084384 [Pa] | $Mw$=0.795368 [g/mol] | $D_1$=7.950774066 [J/mol-K] |
| $\omega$=-0.0479 [-] | $Mw$=1.5473 [g/mol] | $\omega$=-0.2004 [-] | $D_2$=947.6764483 [K] |
| | $\rho_c$=17514.3 [mol/m³] | $V_c$=43.437979 [cm³/mol] | $D_3$=12.96716236 [J/mol-K] |
| | | $\mu_r$=-0.86764340 [-] | $D_4$=346.5603412 [K] |
| | | $\kappa$=-0.00148112 [-] | $D_5$=2734.218834 [J/mol] |
| | | | $D_6$=0.283552887 [J/mol-K] |
| | | | $D_7$=1.926832080 [-] |

Table 3 lists the statistical parameters of each method used in this study to predict experimental density, viscosity, and thermal conductivity. The density from PR-EoS with optimized parameters was used in the optima calculations. In the original consideration, the recommended values by DIPPR801 [42] were used.

**Table 3.** Results of different parameters in the calculation of properties of pure $H_2$.

| Properties | Source of constants | AAD% | MAD% | $R^2$ | PAP |
|---|---|---|---|---|---|
| Density ($\rho$) | Optimal | 0.182 | 1.90 | 0.9993 | 99.85 |
|  | Original | 2.759 | 8.05 | 0.9859 | 94.01 |
| Viscosity ($\mu$) | Optimal | 4.082 | 13.0 | 0.9849 | 96.50 |
|  | Original | 11.82 | 54.3 | 0.8830 | 77.26 |
| Thermal conductivity ($\lambda$) | Optimal | 2.140 | 12.1 | 0.9945 | 98.39 |
|  | Original | 36.01 | 44.0 | 0.9677 | Null |

Note: Definitions of statistical parameters can be found elsewhere [44].

Figure 1 shows the scaled results of optimal and original methods in a cross-plot. Figure 1, alongside Table 3, clearly illustrate how these minor optimizations of inputs drastically affect the outputs. Current optimization has sensible effects on thermal conductivity, viscosity, and density.

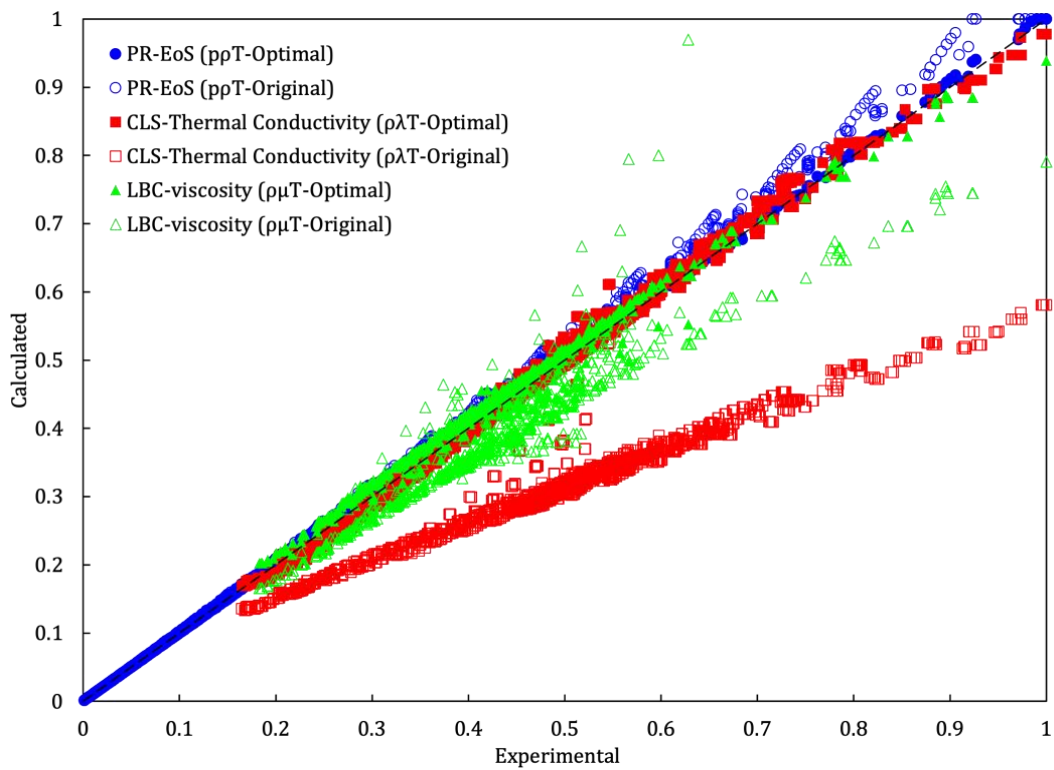

**Figure 1.** Cross-plot of the scaled results of optimal and original methods.

A comprehensive database of thermodynamic and transport properties for various fluids is provided in REFPROP 10 [43], which covers pure substances and mixtures. A total of 38613 data points were generated by REFPROP 10 [43] within the temperature range of 90-1000 K and pressure range of 0.1 to 100 MPa to demonstrate the effectiveness of the optimization process. Table 4 compares the results of models with original and optimal against these 38613 data points. One of the few available experimental measurements of volumetric properties of $H_2$ in open literature is the work conducted by Knapp and colleagues [45]. Figure 2 shows the comparison of the current optimization (solid lines) and original values (dash lines) against experimental data from Knapp et al. [45] for estimating the isobaric heat capacity of $H_2$.

**Table 4.** Statistical metrics of the optimization in this study against original values compared to REFPROP [43].

| | AAD% | | MAD% | | $R^2$ | | PAP | |
|---|---|---|---|---|---|---|---|---|
| | Original | Optimal | Original | Optimal | Original | Optimal | Original | Optimal |
| $\rho$ | 2.487 | 0.274 | 8.343 | 3.930 | 0.99440 | 0.99925 | 97.09 | 99.76 |
| $\mu$ | 14.18 | 2.088 | 175.3 | 11.96 | 0.62749 | 0.99774 | 62.85 | 98.42 |
| $\lambda$ | 38.96 | 1.407 | 45.66 | 11.28 | 0.94011 | 0.99639 | Null | 98.94 |
| $\varphi$ | 5.622 | 0.175 | 33.59 | 5.95 | 0.96900 | 0.99920 | 82.69 | 99.84 |
| $U$ | 7.563 | 0.407 | 28.84 | 13.13 | 0.99963 | 0.99991 | 94.62 | 99.71 |
| $H$ | 3.561 | 0.288 | 14.94 | 5.906 | 0.99969 | 0.99996 | 97.47 | 99.79 |
| $S$ | 107.6 | 2.190 | 438.3 | 53.58 | 0.76736 | 0.99399 | Null | 98.37 |
| $C_v$ | 3.134 | 1.797 | 42.84 | 10.44 | 0.68073 | 0.91356 | 64.63 | 88.69 |
| $C_p$ | 1.231 | 0.901 | 26.62 | 9.445 | 0.10598 | 0.85016 | 31.75 | 88.84 |
| $w$ | 1.268 | 0.884 | 14.49 | 15.18 | 0.99525 | 0.99633 | 98.27 | 98.92 |
| $\mu_{JT}$ | 58.89 | 13.99 | 554714 | 35413 | 0.80543 | 0.97316 | 26.55 | 89.87 |
| $\kappa_T$ | 2.517 | 1.585 | 24.95 | 15.00 | 0.99999 | 0.99999 | 98.22 | 98.87 |
| $\alpha_p$ | 3.288 | 3.058 | 12.29 | 5.781 | 0.99919 | 0.99634 | 97.65 | 97.81 |

Note: absolute average deviation (AAD), maximum absolute deviation (MAD), coefficient of determination ($R^2$), and the percentage of accuracy-precision (PAP), density ($\rho$), viscosity ($\mu$), thermal conductivity ($\lambda$), fugacity coefficient ($\varphi$), internal energy ($U$), enthalpy ($H$), entropy ($S$), isochoric heat capacity ($C_v$), isobaric heat capacity ($C_p$), speed of sound ($w$), Joule-Thomson coefficient ($\mu_{JT}$), isothermal compressibility ($\kappa_T$), and coefficient of thermal expansion ($\alpha_p$).

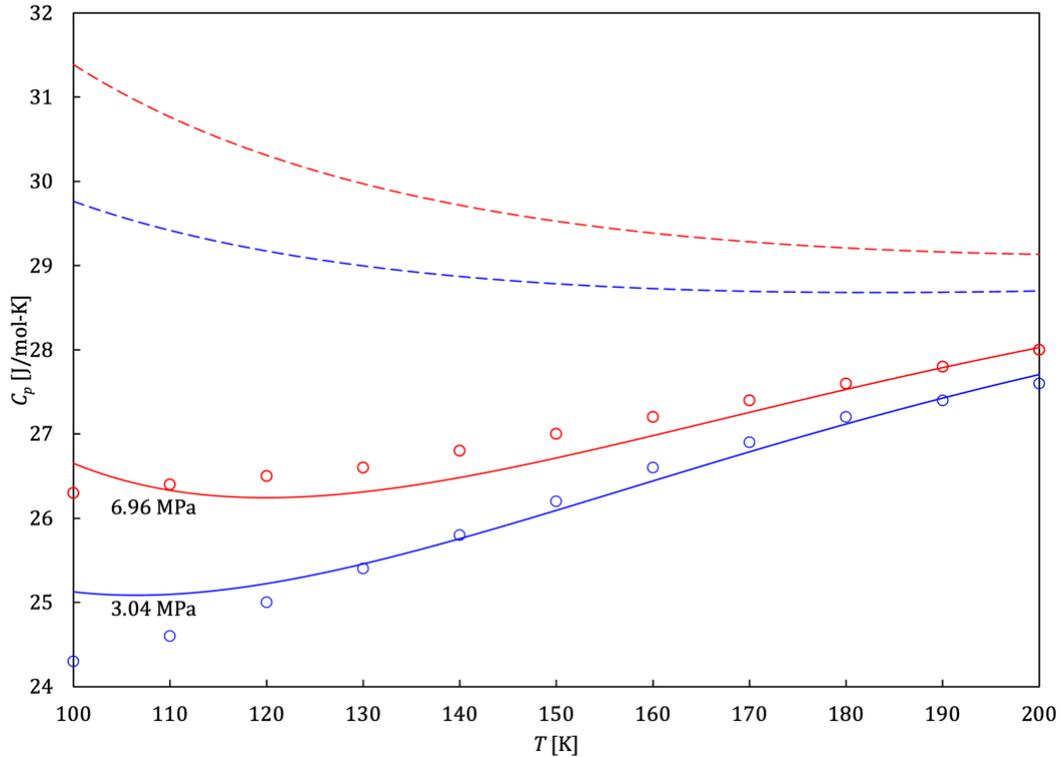

**Figure 2.** Comparison of the current optimization (solid lines) and original values (dash lines) against experimental data from Knapp et al. [45] for estimating the isobaric heat capacity of $H_2$.

## 4. Binary mixtures with hydrogen

### 4.1. Phase equilibria and volumetric properties

#### 4.1.1. Critical points

On the one hand, CEoSs, including PR-EoS, are obtained to do phase equilibria calculations, but in applications, they are used to calculate thermodynamic properties. On the other hand, all EoSs are developed to model pure fluids. However, they extended to the mixtures by the mixing rules. As the nature of fluids differs, the use of binary interaction parameters (BIP or $k_{ij}$) is vital. Conventionally, $k_{ij}$ is used to describe vapor-liquid equilibrium (VLE), and bubble- and dew-point data are used in its determination. But in the context of $H_2$ usage (and this study), it is almost in a gaseous phase. As illustrated by Jarrahian and colleagues [46—48], the critical point of gas mixtures influences the reliability of all thermophysical properties.

     Here, we address this issue by adjusting interaction parameters with respect to the critical point of binary mixtures [49-52], applying Taylor series expansion of Helmholtz free energy ($A$), and considering the second and third order terms [53] for a mixture of total components of N must satisfy

$$Q\Delta \mathbf{N} = 0, \qquad \Delta \mathbf{N}^T \Delta \mathbf{N} = 1 \tag{15}$$

and

$$C = \sum_i \sum_j \sum_k \Delta N_i \Delta N_j \Delta N_k \left( \frac{\partial^3 A}{\partial N_i \partial N_j \partial N_k} \right)_{T,V} = 0 \tag{16}$$

and

$$Q_{ij} = \left( \frac{\partial^2 A}{\partial N_i \partial N_j} \right)_{T,V} = RT \left( \frac{\partial \ln f_i}{\partial N_j} \right)_{T,V} \tag{17}$$

Nested iterations are needed to evaluate critical points. At a fixed volume value, the Newton iteration determines a temperature where the set of homogeneous equations (15) has a nontrivial solution. The element $\Delta N_i$ is calculated to correct volume in an outer loop, and the cubic form $C$ is evaluated. For the mixtures containing $H_2$ at the gaseous region (T > 90 K), values of $k_{ij}$ are listed in Table 5. Figure 3 shows the result of using calculated $k_{ij}$ of this study in estimating critical loci while they have been juxtaposed with values of GERG-2008 [54] multiparameter EoS. In this figure, filled symbols are critical points of less volatile compounds, and the unfilled symbols were obtained by extrapolating saturated liquid and vapor data points. The type III phase behavior, according to the classification scheme of van Konynenburg and Scott [55], is evident in this figure.

**Table 5.** Values of $k_{ij}$ for mixtures containing $H_2$ at T>90 K.

| Mixture | $H_2$-$N_2$ | $H_2$-CO | $H_2$-$CH_4$ | $H_2$-$CO_2$ |
|---|---|---|---|---|
| $k_{ij}$ | -0.13563 | -0.19305 | -0.33516 | -0.83101 |

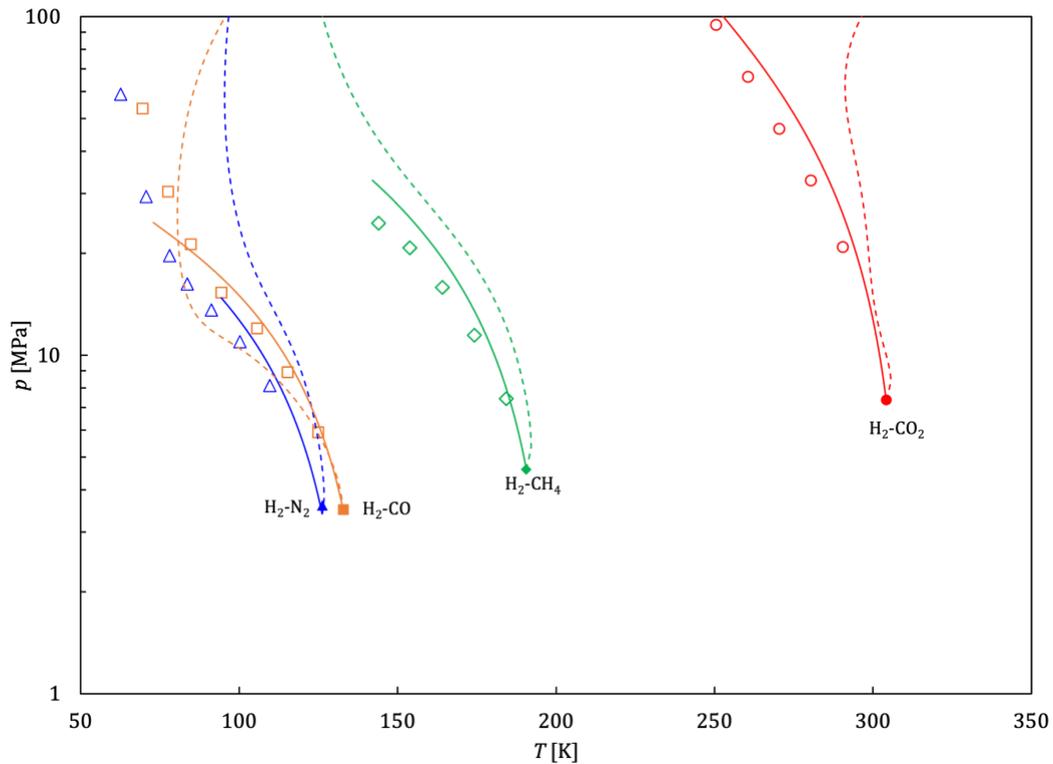

**Figure 3.** p-T diagram of critical loci of the $H_2$ binary mixtures containing $N_2$, CO, $CH_4$, and $CO_2$. Filled symbols are critical points of pure, less volatile compounds, and unfilled symbols are critical points of binary mixtures [49-52]. Solid lines are calculated using PR-EoS (amendments and $k_{ij}$ of this study), and dash lines are the results of GERG-2008 [54].

All the values presented in Table 5 are negative. They exhibit a linear relationship when plotted against the van der Waals volume ($V_W$) [56]. This suggests that as the size of less volatile compounds increases, their interaction with hydrogen also increases. Intuitively, it can be inferred that the equation remains reliable for molecules with a van der Waals volume of 15.116 to 20.668 m³/mole.

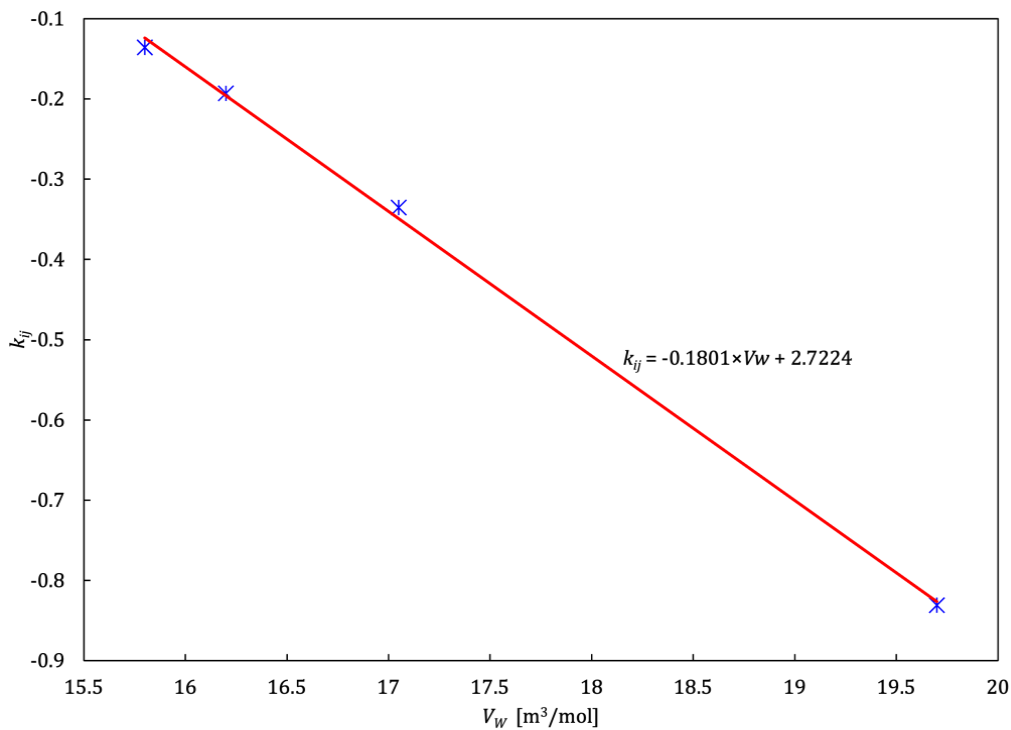

**Figure 4.** Correlation of $k_{ij}$ for critical mixtures as a function of van der Waals volume of less volatile compound.

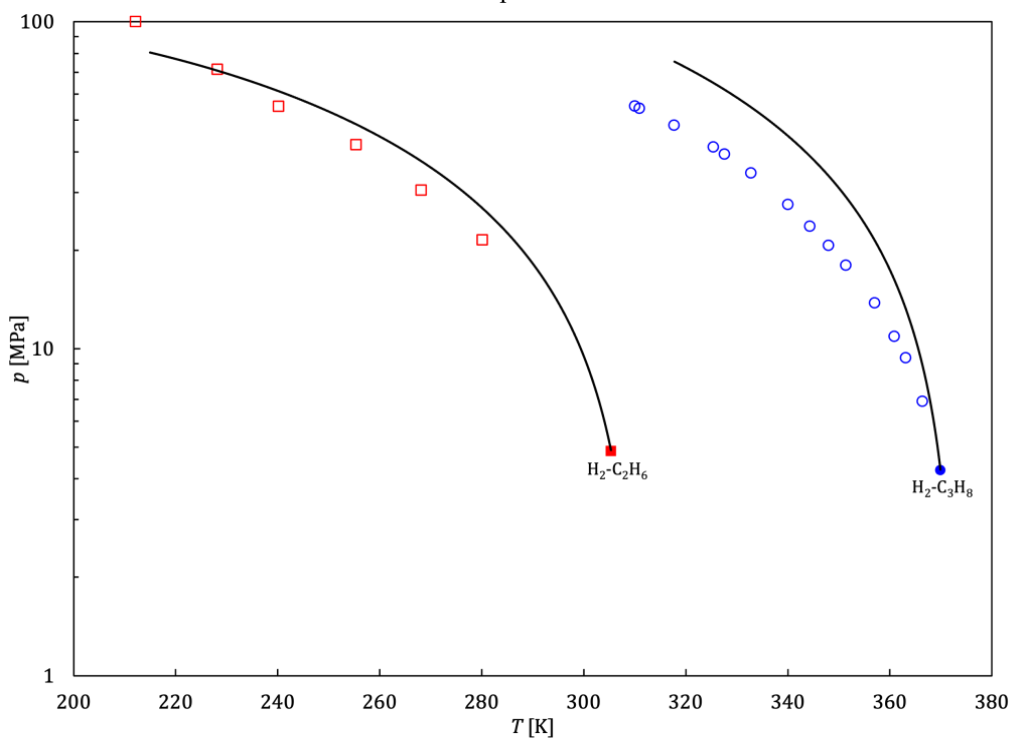

**Figure 5.** p-T diagram of critical loci of the $H_2$ binary mixtures containing $C_2H_6$ and $C_3H_8$.

Investigating a single phase may not be as relevant for systems with larger molecules as the presence of heavier compounds in streams dealing with SMR is uncommon. Still, it is reasonable to consider at least two more hydrocarbons, namely ethane and propane. Figure 5 shows the result of using the proposed correlation in determining critical loci of the H$_2$ binary mixtures containing C$_2$H$_6$ [XX] and C$_3$H$_8$ [YY]. While GERG-2008 [54] failed to locate critical points of these mixtures, the current simple method has reliable results.

$$OF = \frac{1}{N}\sum_{i=1}^{N}\left(\left(x_{1,i} - x_{1,i,calc}\right)2 + \left(y_{1,i} - y_{1,i,calc}\right)^2\right) \qquad (15)$$

where N is the number of state points included in the regression, $x_{1,i}$ is the mole fraction of component 1 (the more volatile component, which here is H$_2$) at point $i$ from experimental measurements, and $x_{1,i,calc}$ is the mole fraction at that point calculated by the PR-EoS [21].

## 5. Conslusions

The work introduces several amendments to widely used semi-analytical methods, including the Peng-Robinson Equation of State (PR-EoS), Jossi-Stiel-Thodos/Lohrenz-Bray-Clark (JST/LBC) viscosity correlation, and Chung-Lee-Starling (CLS) thermal conductivity correlation. These amendments are designed to improve the accuracy of calculations for pure hydrogen and its binary mixtures with methane, nitrogen, carbon monoxide, and carbon dioxide. The modifications leave the mathematical structure of these models intact to maintain compatibility with commercial software.

The study covers temperatures ranging from 90 K to 1000 K and pressures from atmospheric up to 100 MPa. The introduced amendments reduce the Absolute Average Deviation (AAD) for hydrogen density from 2.76% to 0.182%, viscosity from 11.82% to 4.08%, and thermal conductivity from 36% to 2.14%. The study also determines binary interaction parameters for hydrogen mixtures, improving the calculation of transport properties and phase equilibria. Parameters were optimized using a database of experimental data and the Nelder-Mead simplex method. Results are benchmarked against both experimental data and existing models, such as those in REFPROP 10. Statistical analysis shows significant improvement in property predictions (density, viscosity, thermal conductivity) when compared to original models. The work focuses on hydrogen's role in energy applications, particularly hydrogen production, storage, and transportation, where accurate predictions of thermophysical properties are essential for infrastructure design, safety, and efficiency.

The proposed amendments improve the accuracy of widely accepted models for calculating hydrogen and hydrogen mixture properties. These enhancements will enable more precise modeling of hydrogen storage and transportation systems, thereby supporting the transition to hydrogen as a key energy carrier. The optimized models provide better predictions for critical industrial applications, such as pipeline transport and underground hydrogen storage, and help meet future energy needs with higher efficiency and safety.

## Acknowledgements

This work was supported by the School of Energy Resources Hydrogen Research Center (SER H2ERC) at the University of Wyoming. The corresponding author thanks Director Eugene Holubnyak for our lively discussions.

## Conflict of interest

The authors declare no competing interest.

## References

[1] Yadav, D., Lu, X., Vishwakarma, C. B., & Jing, D. (2023). Advancements in microreactor technology for hydrogen production via steam reforming: A comprehensive review of experimental studies. Journal of Power Sources, 585, 233621. https://doi.org/10.1016/j.jpowsour.2023.233621

[2] Ban, D., & Bebić, J. (2023). An introduction of future fuels on working ship for GHGs reduction: Trailing suction hopper dredger case study. Journal of Cleaner Production, 405, 137008. https://doi.org/10.1016/j.jclepro.2023.137008

[3] Yin, K., Wei, R., Ruan, J., Cui, P., Zhu, Z., Wang, Y., & Zhao, X. (2023). Life cycle assessment and life cycle cost analysis of surgical mask from production to recycling into hydrogen process. Energy, 129225. https://doi.org/10.1016/j.energy.2023.129225

[4] Müller, M., Pfeifer, M., Holtz, D., & Müller, K. (2024). Comparison of green ammonia and green hydrogen pathways in terms of energy efficiency. Fuel, 357, 129843. https://doi.org/10.1016/j.fuel.2023.129843

[5] Koo, B., Ha, Y., & Kwon, H. (2023). Preliminary evaluation of hydrogen blending into high-pressure natural gas pipelines through hydraulic analysis. Energy, 268, 126639. https://doi.org/10.1016/j.energy.2023.126639

[6] Nova, A., Prifti, K., Negri, F., & Manenti, F. (2023). Multiscale techno-economic analysis of orange hydrogen synthesis. Energy, 282, 128644. https://doi.org/10.1016/j.energy.2023.128644

[7] Liu, X., Michal, G., Godbole, A., & Lu, C. (2021). Decompression modelling of natural gas-hydrogen mixtures using the Peng-Robinson equation of state. International Journal of Hydrogen Energy, 46(29), 15793-15806. https://doi.org/10.1016/j.ijhydene.2021.02.129

[8] Tan, K., Mahajan, D., & Venkatesh, T. A. (2023). Computational fluid dynamic modeling of methane-hydrogen mixture transportation in pipelines: Understanding the effects of pipe roughness, pipe diameter and pipe bends. International Journal of Hydrogen Energy. https://doi.org/10.1016/j.ijhydene.2023.06.195

[9] Jia, W., Ren, Q., Zhang, H., Yang, M., Wu, X., & Li, C. (2023). Multicomponent leakage and diffusion simulation of natural gas/hydrogen mixtures in compressor plants. Safety Science, 157, 105916. https://doi.org/10.1016/j.ssci.2022.105916

[10] Li, H., Cao, X., Du, H., Teng, L., Shao, Y., & Bian, J. (2022). Numerical simulation of leakage and diffusion distribution of natural gas and hydrogen mixtures in a closed container. International Journal of Hydrogen Energy, 47(84), 35928-35939. https://doi.org/10.1016/j.ijhydene.2022.08.142

[11] De Bellis, V., Malfi, E., Bozza, F., Cafari, A., Caputo, G., Yvonnen, J., Di Miceli, A., Leino J., Grahn, V. (2023). A thermodynamic model of a rapid compression expansion machine applied for the estimation of the flame speed of air/hydrogen mixtures. Fuel, 341, 127657. https://doi.org/10.1016/j.fuel.2023.127657

[12] Chen, Z., Ji, Y., Zhang, H., Guo, L., Chen, L., Li, X., & Li, F. (2023). Numerical study on the flow and heat transfer characteristics in a high-temperature gas-cooled reactor core. Nuclear Engineering and Design, 413, 112544. https://doi.org/10.1016/j.nucengdes.2023.112544

[13] Fang, Y., Yu, Q., Wang, C., Tian, W., Su, G., & Qiu, S. (2023). Heat transfer of hydrogen with variable properties in a heated tube. International Journal of Heat and Mass Transfer, 209, 124128. https://doi.org/10.1016/j.ijheatmasstransfer.2023.124128

[14] Naquash, A., Qyyum, M. A., Chaniago, Y. D., Riaz, A., Yehia, F., Lim, H., & Lee, M. (2023). Separation and purification of syngas-derived hydrogen: A comparative evaluation of membrane-and cryogenic-assisted approaches. Chemosphere, 313, 137420. https://doi.org/10.1016/j.chemosphere.2022.137420

[15] Lee, W., Kang, D. W., Ahn, Y. H., & Lee, J. W. (2023). Blended hydrate seed and liquid promoter for the acceleration of hydrogen hydrate formation. Renewable and Sustainable Energy Reviews, 177, 113217. https://doi.org/10.1016/j.rser.2023.113217

[16] Li, J. Q., Chen, Y., Ma, Y. B., Kwon, J. T., Xu, H., & Li, J. C. (2023). A study on the Joule-Thomson effect of during filling hydrogen in high pressure tank. Case Studies in Thermal Engineering, 41, 102678. https://doi.org/10.1016/j.csite.2022.102678

[17] Chai, M., Chen, Z., Nourozieh, H., & Yang, M. (2023). Numerical simulation of large-scale seasonal hydrogen storage in an anticline aquifer: A case study capturing hydrogen interactions and cushion gas injection. Applied Energy, 334, 120655. https://doi.org/10.1016/j.apenergy.2023.120655

[18] Shoushtari, S., Namdar, H., & Jafari, A. (2023). Utilization of $CO_2$ and $N_2$ as cushion gas in underground gas storage process: A review. Journal of Energy Storage, 67, 107596. https://doi.org/10.1016/j.est.2023.107596

[19] Al-Mahgari, A. M., Moh'd A, A. N., & Khashan, S. A. (2023). Pressurized green hydrogen from water electrolysis: Compression before or after electrolysis? A comparison among different configurations. Journal of Energy Storage, 73, 109251. https://doi.org/10.1016/j.est.2023.109251

[20] Hematpur, H., Abdollahi, R., Rostami, S., Haghighi, M., & Blunt, M. J. (2023). Review of underground hydrogen storage: Concepts and challenges. Advances in Geo-Energy Research, 7(2), 111-131. https://doi.org/10.46690/ager.2023.02.05

[21] Peng, D. Y., & Robinson, D. B. (1976). A new two-constant equation of state. Industrial & Engineering Chemistry Fundamentals, 15(1), 59-64. https://doi.org/10.1021/i160057a011

[22] Jossi, J. A., Stiel, L. I., & Thodos, G. (1962). The viscosity of pure substances in the dense gaseous and liquid phases. AIChE Journal, 8(1), 59-63. https://doi.org/10.1002/aic.690080116

[23] Lohrenz, J., Bray, B. G., & Clark, C. R. (1964). Calculating viscosities of reservoir fluids from their compositions. Journal of Petroleum Technology, 16(10), 1171-1176. https://doi.org/10.2118/915-PA

[24] Chung, T. H., Ajlan, M., Lee, L. L., & Starling, K. E. (1988). Generalized multiparameter correlation for nonpolar and polar fluid transport properties. Industrial & engineering chemistry research, 27(4), 671-679. https://doi.org/10.1021/ie00076a024

[25] Elliott J. R., Vladimir D., Knotts IV, Thomas A., Vincent Wilding W. (2023). Properties of gases and liquids. McGraw Hill; 6$^{th}$ edition.

[26] Leachman, J. W., Jacobsen, R. T., Penoncello, S. G., & Lemmon, E. W. (2009). Fundamental equations of state for parahydrogen, normal hydrogen, and orthohydrogen. Journal of Physical and Chemical Reference Data, 38(3), 721-748. https://doi.org/10.1063/1.3160306

[27] Muzny, C. D., Huber, M. L., & Kazakov, A. F. (2013). Correlation for the viscosity of normal hydrogen obtained from symbolic regression. Journal of Chemical & Engineering Data, 58(4), 969-979. https://doi.org/10.1021/je301273j

[28] Assael, M. J., Huber, M. L., Perkins, R. A., & Takata, Y. (2011). Correlation of the Thermal Conductivity of Normal and Parahydrogen from the Triple Point to 1000 K and up to 100 MPa. Journal of Physical and Chemical Reference Data, 40(3). https://doi.org/10.1063/1.3606499

[29] Stiel, L. I., & Thodos, G. (1961). The viscosity of nonpolar gases at normal pressures. AIChE Journal, 7(4), 611-615. https://doi.org/10.1002/aic.690070416

[30] Herning, F., & Zipperer, L. (1936). Calculation of the viscosity of technical gas mixtures from the viscosity of the individual gases. Gas und Wasserfach, 79, 69.

[31] Chung, T. H., Lee, L. L., & Starling, K. E. (1984). Applications of kinetic gas theories and multiparameter correlation for prediction of dilute gas viscosity and thermal

conductivity. Industrial & Engineering Chemistry Fundamentals, 23(1), 8-13. https://doi.org/10.1021/i100013a002

[32] Lee, L. L., Mo, K. C., & Starling, K. E. (1977). Multiparameter Conformal Solution Theory—Application to Hydrocarbon Pure Fluid and Mixture Thermodynamic Properties. Berichte der Bunsengesellschaft für physikalische Chemie, 81(10), 1044-1046. https://doi.org/10.1002/bbpc.19770811035

[33] Bartolomeu, R. A., & Franco, L. F. (2020). Thermophysical properties of supercritical $H_2$ from Molecular Dynamics simulations. International Journal of Hydrogen Energy, 45(33), 16372-16380. https://doi.org/10.1016/j.ijhydene.2020.04.164

[34] Aly, F. A., & Lee, L. L. (1981). Self-consistent equations for calculating the ideal gas heat capacity, enthalpy, and entropy. Fluid Phase Equilibria, 6(3-4), 169-179. https://doi.org/10.1016/0378-3812(81)85002-9

[35] Heidaryan, E., & Aryana, S. A. (2024). Empirical correlations for density, viscosity, and thermal conductivity of pure gaseous hydrogen. Advances in Geo-Energy Research, 11(1), 54-73. https://doi.org/10.46690/ager.2024.01.06

[36] Schlumberger (2023) FluidModeler. Available from: https://www.software.slb.com/products/fluidmodeler

[37] Computer Modelling Group Ltd. (2023) WinProp. Available from: https://www.cmgl.ca/winprop

[38] KBC Advanced Technologies. (2023) Multiflash. Available from: https://www.kbc.global/multiflash-simulation-software

[39] Calsep (2023) PVTsim Nova 6. Available from: https://www.calsep.com/

[40] Aspen Technology (2023) Aspen Plus 14. Available from: https://www.aspentech.com/en/products/engineering/aspen-plus

[41] Nelder, J. A., & Mead, R. (1965). A simplex method for function minimization. The computer journal, 7(4), 308-313. https://doi.org/10.1093/comjnl/7.4.308

[42] Design Institute for Physical Properties, Sponsored by AIChE. (2005; 2008; 2009; 2010; 2011; 2012; 2015; 2016; 2017; 2018; 2019; 2020; 2021). DIPPR Project 801 - Full Version. Design Institute for Physical Property Research/AIChE. https://app.knovel.com/hotlink/toc/id:kpDIPPRPF7/dippr-project-801-full/dippr-project-801-full

[43] Lemmon, E. W., Bell, I. H., Huber, M. L., & McLinden, M. O. (2018). NIST Standard Reference Database 23: Reference Fluid Thermodynamic and Transport Properties REFPROP, Version 10.0. National Institute of Standards and Technology, Standard Reference Data Program. https://doi.org/10.18434/T4/1502528

[44] Heidaryan, E. (2019). A note on model selection based on the percentage of accuracy-precision. Journal of Energy Resources Technology, 141(4), 045501. https://doi.org/10.1115/1.4041844

[45] Knapp, H., Schmölling, K., & Neumann, A. (1976). Measurement of the molal heat capacity of $H_2$-$N_2$ mixtures. Cryogenics, 16(4), 231-237. https://doi.org/10.1016/0011-2275(76)90266-6

[46] Jarrahian, A., & Heidaryan, E. (2014). A new cubic equation of state for sweet and sour natural gases even when composition is unknown. Fuel, 134, 333-342. https://doi.org/10.1016/j.fuel.2014.05.066

[47] Jarrahian, A., Aghel, B., & Heidaryan, E. (2015). On the viscosity of natural gas. Fuel, 150, 609-618. https://doi.org/10.1016/j.fuel.2015.02.049

[48] Jarrahian, A., & Heidaryan, E. (2014). A simple correlation to estimate natural gas thermal conductivity. Journal of Natural Gas Science and Engineering, 18, 446-450. https://doi.org/10.1016/j.jngse.2014.04.005

[49] Tsang, C. Y., et al. "Phase equilibria in the $H_2$/$CH_4$ system at temperatures from 92.3 to 180.0 K and pressures to 140 MPa." Chemical Engineering Communications 6.6 (1980): 365-383. https://doi.org/10.1080/00986448008912543

[50] Streett, W. B., & Calado, J. C. G. (1978). Liquid-vapour equilibrium for hydrogen+ nitrogen at temperatures from 63 to 110 K and pressures to 57 MPa. The Journal of Chemical Thermodynamics, 10(11), 1089-1100. https://doi.org/10.1016/0021-9614(78)90083-6

[51] Tsang, C. Y., & Streett, W. B. (1981). Phase equilibria in the $H_2$- CO system at temperatures from 70 to 125 K and pressures to 53 MPa. Fluid Phase Equilibria, 6(3-4), 261-273. https://doi.org/10.1016/0378-3812(81)85008-X

[52] Tsang, C. Y., & Street, W. B. (1981). Phase equilibria in the $H_2$/$CO_2$ system at temperatures from 220 to 290 K and pressures to 172 MPa. Chemical Engineering Science, 36(6), 993-1000. https://doi.org/10.1016/0009-2509(81)80085-1

[53] Heidemann, R. A., & Khalil, A. M. (1980). The calculation of critical points. AIChE journal, 26(5), 769-779. https://doi.org/10.1002/aic.690260510

[54] Kunz, O., & Wagner, W. (2012). The GERG-2008 wide-range equation of state for natural gases and other mixtures: an expansion of GERG-2004. Journal of chemical & engineering data, 57(11), 3032-3091. https://doi.org/10.1021/je300655b

[55] Van Konynenburg, P. H., & Scott, R. L. (1980). Critical lines and phase equilibria in binary van der Waals mixtures. Philosophical Transactions of the Royal Society of London. Series A, Mathematical and Physical Sciences, 298(1442), 495-540. https://doi.org/10.1098/rsta.1980.0266

[56] Bondi, A. V. (1964). van der Waals volumes and radii. The Journal of physical chemistry, 68(3), 441-451. https://doi.org/10.1021/j100785a001

[57] Le Guennec, Y., Lasala, S., Privat, R., & Jaubert, J. N. (2016). A consistency test for α-functions of cubic equations of state. Fluid Phase Equilibria, 427, 513-538. https://doi.org/10.1016/j.fluid.2016.07.026

[58] Sun, X., Fang, Y., Zhao, W., & Xiang, S. (2022). New Alpha Functions for the Peng–Robinson Cubic Equation of State. ACS omega, 7(6), 5332-5339. https://doi.org/10.1021/acsomega.1c06519

[59] Jaubert, J. N., Qian, J., Privat, R., & Leibovici, C. F. (2013). Reliability of the correlation allowing the $k_{ij}$ to switch from an alpha function to another one in hydrogen-containing systems. Fluid Phase Equilibria, 338, 23-29. https://doi.org/10.1016/j.fluid.2012.10.016

[60] Qian, J. W., Jaubert, J. N., & Privat, R. (2013). Phase equilibria in hydrogen-containing binary systems modeled with the Peng–Robinson equation of state and temperature-dependent binary interaction parameters calculated through a group-contribution method. The Journal of Supercritical Fluids, 75, 58-71. https://doi.org/10.1016/j.supflu.2012.12.014